\documentclass[sn-mathphys,Numbered]{sn-jnl}
\usepackage{graphicx}%
\usepackage{multirow}%
\usepackage{amsmath,amssymb,amsfonts}%
\usepackage{amsthm}%
\usepackage{mathrsfs}%
\usepackage[title]{appendix}%
\usepackage{textcomp}%
\usepackage{manyfoot}%
\usepackage{booktabs}%
\usepackage{listings}%

	\numberwithin{equation}{section} %number equations by section
\usepackage[table]{xcolor}
\usepackage{enumitem}[shortlabels]
\usepackage{thm-restate}
\usepackage{hyperref}       % hyperlinks
\usepackage{cleveref}       % for not having to write Proposition~\ref{}. Write \Cref{label} for "Theorem X" and \cref{label} for "theorem X"
\usepackage{nicefrac}       % compact symbols for 1/2, etc.
\usepackage{microtype}      % microtypography	
\usepackage{graphbox}       % for vertical alignment of multiple images in figures
\usepackage{makecell}       % for two-line cells

\usepackage{ifthen}
\usepackage[caption=false]{subfig}
\newboolean{commentsactivated}
\setboolean{commentsactivated}{false}
\newcommand{\co}[1]{\ifthenelse{\boolean{commentsactivated}}{{\color{teal} {\em CO: #1 }}}{}}
\newcommand{\vc}[1]{\ifthenelse{\boolean{commentsactivated}}{{\color{blue} {\em VC: #1 }}}{}}
\newcommand{\vk}[1]{\ifthenelse{\boolean{commentsactivated}}{{\color{red} {\em VK: #1 }}}{}}
\newcommand{\cs}[1]{\ifthenelse{\boolean{commentsactivated}}{{\color{cyan} {\em CS: #1 }}}{}}
\newcommand{\cvm}[1]{\ifthenelse{\boolean{commentsactivated}}{{\color{brown} {\em CM: #1 }}}{}}

\newcommand{\im}[1]{\ifthenelse{\boolean{commentsactivated}}{{\color{pink} {\em IM: #1 }}}{}}

\newcommand{\draftOnly}[1]{\ifthenelse{\boolean{commentsactivated}}{#1}{}}

\newboolean{highlightEdits}
\setboolean{highlightEdits}{false}
\newcommand{\edit}[1]{\ifthenelse{\boolean{highlightEdits}}{{\color{purple}{#1}}}{#1}}

\newboolean{journalTextShown}
\setboolean{journalTextShown}{false}
\newcommand{\journal}[1]{\ifthenelse{\boolean{journalTextShown}}{{{\ifthenelse{\boolean{highlightEdits}}{{\color{olive}{#1}}}{#1}}}}{}}
\newboolean{anonymous}
\setboolean{anonymous}{false}
\newcommand{\anonymise}[2]{\ifthenelse{\boolean{anonymous}}{{{#2}}}{#1}}
\newtheorem{definition}{\protect\definitionname}
\theoremstyle{definition}           % Non-italic environments

\providecommand{\corollaryname}{Corollary}
\providecommand{\claimname}{Claim}
\providecommand{\definitionname}{Definition}
\providecommand{\lemmaname}{Lemma}
\providecommand{\notationname}{Notation}
\providecommand{\remarkname}{Remark}
\providecommand{\problemname}{Problem}
\providecommand{\propositionname}{Proposition}
\providecommand{\examplename}{Example}
\providecommand{\theoremname}{Theorem}
\providecommand{\conjecturename}{Conjecture}
\providecommand{\observationname}{Observation}

\DeclareMathAlphabet{\mathpzc}{OT1}{pzc}{m}{it}
\DeclareMathSymbol{\shortminus}{\mathbin}{AMSa}{"39}

\DeclareMathSymbol{\shortminus}{\mathbin}{AMSa}{"39}

\usepackage{pifont}

\begin{document}

\title[AI X-Risk: Invisible to Science?]{\centering Extinction Risks from AI: Invisible to Science?}

% \author{\fnm{Anonymous} \sur{Author(s)}}

\author*[1]{\fnm{Vojtech} \sur{Kovarik}}\email{vojta.kovarik@gmail.com}

\author[1]{\fnm{Christian} \sur{van Merwijk}}\email{cpv@andrew.cmu.edu}

\author[2]{\fnm{Ida} \sur{Mattsson}}\email{imattsso@andrew.cmu.edu}


\affil*[1]{\orgdiv{Foundations of Cooperative AI Lab, Computer Science Department}, \orgname{Carnegie Mellon University}, \orgaddress{\street{5000 Forbes Avenue}, \city{Pittsburgh}, \postcode{15 213}, \state{PA}, \country{United States}}}

\affil[2]{\orgdiv{Department of Philosophy}, \orgname{Carnegie Mellon University}, \orgaddress{\street{5000 Forbes Avenue}, \city{Pittsburgh}, \postcode{15 213}, \state{PA}, \country{United States}}}

% \affil[2]{\orgdiv{Department}, \orgname{Organization}, \orgaddress{\street{Street}, \city{City}, \postcode{10587}, \state{State}, \country{Country}}}

% \affil[3]{\orgdiv{Department}, \orgname{Organization}, \orgaddress{\street{Street}, \city{City}, \postcode{610101}, \state{State}, \country{Country}}}

\abstract{In an effort to inform the discussion surrounding existential risks from AI,
we formulate Extinction-level Goodhart’s Law as
``\textit{Virtually any goal specification, pursued to the extreme, will result in the extinction of humanity}'',
and we aim to understand which formal models are suitable for investigating this hypothesis.
Note that we remain agnostic as to whether Extinction-level Goodhart's Law holds or not.
As our key contribution, we identify a set of conditions that are necessary for a model that aims to be informative for evaluating specific arguments for Extinction-level Goodhart's Law.
Since each of the conditions seems to significantly contribute to the complexity of the resulting model, formally evaluating the hypothesis might be exceedingly difficult.
This raises the possibility that whether the risk of extinction from artificial intelligence is real or not,
    the underlying dynamics might be invisible to current scientific methods.

}




\maketitle

\section{Introduction}\label{sec:intro}

% Explain how this paper is motivated by AI X-risk discussions.
There is an important ongoing debate regarding the possibility that artificial intelligence (AI) might cause either the literal extinction of humanity or other comparably undesirable outcomes \cite{XriskStatement}.
For the purpose of this paper, we take no stance on whether this possibility will materialise.
However, we believe a valuable contribution is to investigate the tools by which we can promote an informed discussion.

An unfortunate aspect of the hypothesis that AI might cause human extinction is that
    unless we are already confident the hypothesis is false,
    it is one that we should not test empirically.
Moreover,
    even less direct ways of obtaining empirical evidence about AI risk typically require building powerful AI,
    and therefore increase the extinction risk, if such a risk indeed exists.
For this reason, it is crucial to also search for theoretical means of advancing the discussion of AI risk.

% Several mechanisms with detrimental outcomes have been proposed, including the misuse of AI, systemic risks from AI (such as human disempowerment), and accident risk from AI \cite{systemicRiskDafoe}.
While several different mechanisms have been proposed to potentially contribute to AI risk \cite{systemicRiskDafoe},
    this paper only aims to contribute to a narrower discussion,
    of potential extinction risks
        posed by the unintentional development or deployment of a powerful agentic\footnotemark{} AI
        whose goals are misaligned with the intention of its designers.
    \footnotetext{By 'agentic' we refer to characteristics that are inherent in the generic functioning of a specific AI system, rather than deriving from use or interactions within particular environments.}
However, an important part evaluating this type of risk is predicting the most likely capabilities of future AI systems, which is beyond the scope of this paper.
To sidestep this issue,
    this text will discuss AI agents which are \textit{by assumption} capable of arbitrarily powerful optimisation,
    and remain agnostic as to whether such agents are realistic.
    % \co{Not so important, but: One thing I also didn't understand was why you consider the ``in the limit'' / infinite optimization power version of Goodhart's law. Is there actually any argument for which it's necessary to consider this case as opposed to some version of real-world superintelligence?}
    % \vk{(1) That's a good point, maybe it proves more than it claims. (2) I wanted to avoid a discussion of whether \textit{any} superintelligence is possible or not. (3) For the real world thing, I would make a different claim.}

\textbf{An important part of investigating the extinction risk from agentic AI
    is identifying models which are informative for evaluating arguments for such risk.
Our key contribution is a list of five conditions which are necessary for any such model.}
We believe that these findings have important implications for the difficulty of properly assessing the risk posed by AI.
    However, we defer the discussion of these implications to \Cref{sec:conclusion}.
Note that while we focus on a particular argument,
    we believe that the conditions we identify are sufficiently robust to also apply to a range of alternative arguments for the same claim.
We thus believe that our high-level conclusions are not overly sensitive to the choice of the argument.

In the remainder of this section, we present an analogy
    which clarifies what we mean by ``conditions necessary for a model to be informative for evaluating an argument $A$ for a hypothesis $H$''
    and illustrates the central idea used to generate the paper's findings.

\subsection{Illustrative Example}\label{sec:sub:methodology_example}

To illustrate the methodology used in the remainder of the paper, consider the following fictitious scenario where
    the debate about the hypothesis that AI might cause human extinction
    is replaced by
    a debate about the hypothesis that a particular rocket will fail to land on the Moon.

\smallskip
\noindent
    \textbf{Alice:} Look, I have built a rocket. I am sure that if I launch it, it will land on the Moon.\\
    \textbf{Bob:} Actually, I see a thousand arguments why this will fail. For example: %I think the rocket will miss. 
    %\textbf{Alice:} No, if I aim it in this direction, it will definitely hit.\\
    %\textbf{Bob:} Uhm. I see a thousand arguments why this will fail.
    %    But if you want a simple specific argument, consider this one:
        ``\emph{the rocket will miss because \dots
        \begin{itemize}[leftmargin=1.5cm]
            \item[(A1)] The rocket is currently pointing directly at the Moon.
            \item[(A2)] However, the Moon will eventually move away from where it was at the start.
            \item[(A3)] And the rocket is not fast enough to get there before this happens.''
        \end{itemize}}
    \noindent
    \textbf{Alice:}
        Well, that all sounds plausible. %but I still think I am right;
            %but perhaps \im{mod} if we just point it at the moon and make it go fast enough?
        Let us resolve this using a more formal model.\\
    \textbf{Bob:} Sounds good. But which model should we use?

\medskip
One strategy for Alice and Bob is to look for a realistic model that allows them to accurately predict what will \textit{actually} happen with the rocket.
    However, such model might include complicated aspects of physics and mathematics -- for example, differential calculus -- that neither Bob nor Alice understand.
    Even worse, the debate might be taking place in a world where differential calculus has not yet been invented.

A more tractable approach is to choose a model that is inaccurate but remains informative for addressing the intended goal of evaluating the \textit{argument} (A1-3).\footnote{
        Note that 
        Bob's argument cannot rule out the possibility that the rocket will land on the moon for reasons unrelated to Alice's and Bob's discussion.
    }
While there might be a multitude of such models,
    we can immediately tell that any such model must, \textit{necessarily},
    be able to capture the dynamics that are central to Bob's argument.
For example, to evaluate the argument (A1), that the rocket is initially pointing towards to Moon, the model must, at the minimum, satisfy the following condition:
\begin{itemize}[leftmargin=1.5cm]
        \item[(NC1)] The model must allow us to talk about the direction of the Moon and the direction the rocket is pointing.
    \end{itemize}
Similarly, any model that allows for evaluating the arguments (A2) and (A3) must necessarily satisfy the conditions (NC2) and (NC3):
    \begin{itemize}[leftmargin=1.5cm]
        \item[(NC2)] The model must assume that the position of the Moon changes over time.
        \item[(NC3)] The model must allow us to talk about the time it takes the rocket to reach the original position of the Moon.\footnotemark{}
    \end{itemize}
    \footnotetext{
        As we will discuss in detail later, there is one other necessary condition that plays an important role:
        (NC0) \textit{When making the model, we must have at least made an attempt to ensure that the model accurately reflects reality.}
        For example, it would not work to make up a random number and claim that this how fast the Moon moves.
    }

\begin{figure}[h]%
    \centering
    \subfloat{\includegraphics[scale=0.3]{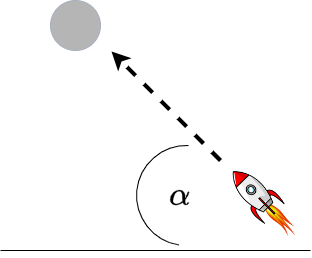}}%
    \qquad
    \subfloat{\includegraphics[scale=0.3]{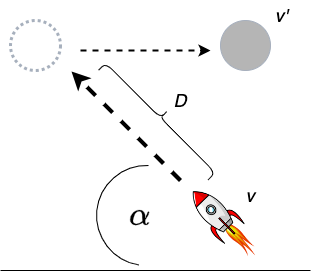}}%
    \caption{Two models for evaluating an argument that ``a rocket will fail to land on the Moon \textit{because the Moon moves}''.
    The model (a) is clearly uninformative for this purpose, since it fails to capture this key dynamic.
    The model (b) is also uninformative, but less obviously so (it is missing the diameter of the Moon).
    Both models are inaccurate, but this is irrelevant for the purpose of evaluating the particular argument that was given.
    }%
    \label{fig:Rockets}%
\end{figure}

Once we have identified the necessary conditions above,
    we can use them to quickly rule out models that are clearly unsuitable for evaluating the argument given by Bob.
For example,
    the model illustrated in \Cref{fig:Rockets}(a) is clearly uninformative for evaluating the argument
    because it does not account for the fact that the Moon moves.
In contrast,
    the conditions (NC1-3) do not rule out the more complex model in \Cref{fig:Rockets}(b).

Admittedly,
    this approach does not ensure that we arrive at a sufficiently informative model,
    and attempting to find an approach that does so would be beyond the scope of this paper.
Nevertheless,
    the approach can help with this goal by making it easier to notice the missing features that need to be added to the model --- such as the diameter of the Moon in the rocket example.

\medskip

To summarise,
    we considered a particular hypothesis $H$ (``the rocket will miss''),
    and focused on a particular argument (A1)-(A3) for $H$.
We then identified a set of necessary conditions (NC1)-(NC3) that must be satisfied by any model that is hoping to be informative towards evaluating the argument.
This allowed us to quickly rule out certain models as uninformative, and gave us a better intuition for the likely properties of informative models.

\subsection{Overview}\label{sec:sub:overview}

As indicated earlier, the key contribution of this paper is
    performing an analysis analogous to \Cref{sec:sub:methodology_example} for (a particular argument for) the hypothesis that pursuing virtually any goal specification to the extreme would cause the extinction of humanity.
The remainder of this text is organised as follows.
    In \Cref{sec:argument}, we describe the aforementioned argument for extinction risk from extremely powerful optimisation.
    In \Cref{sec:properties}, we identify a set of corresponding properties that are necessary for a model to be informative towards evaluating the argument from \Cref{sec:argument}.
    In \Cref{sec:related_work}, we discuss the connections between the present papers and the most closely related concepts and results in the existing literature.
    Finally, we discuss the implications of our findings in \Cref{sec:conclusion}.
    
    To clarify potential misunderstandings regarding the intended interpretation of the arguments or necessary conditions, we include \Cref{sec:additional_content}
        which presents a number of additional illustrative examples.
        This section also shows that none of the necessary conditions we list is redundant.

\section{An Argument for Extinction Risk from Arbitrarily Powerful Optimisation}\label{sec:argument}

In this section, we describe a particular argument for the following hypothesis:

\medskip
\begin{definition}[informal]\label{def:weak-version-CGL}
    The \textbf{Weak Version of Extinction-level Goodhart's Law} is the hypothesis that
    \emph{virtually any goal specification, pursued to the extreme, will result in the extinction of humanity}.
    % \emph{virtually any quantifiable goal, taken towards its absolute maximum, will result in the extinction of humanity}.
\end{definition}

\medskip
Four crucial clarifications are in place:
First,
    the formulation of \Cref{def:weak-version-CGL} does not constrain ``goals'' to rely only on comparison of relative outcomes.\footnote{It is assumed that the goal can be optimized, i.e. is subject to a preference ordering that identifies a unique (set of) actions, but such goals could equally well be defined on the basis on, for example, a deontological commitment.}
 %   \co{Don't know what ``utilitarian goals'' means...}
 %   \vk{potential footnote placeholder: we could expand on this if needed, give an example or such}. 
Second,
    the qualification ``weak version'' is meant to indicate that extinction only happens in the limit of extreme optimisation.
    The claim should thus be viewed as importantly different from
        the prediction that optimisation \textit{will}, in the future, result in the extinction of humanity,
        or a ``quantitative version'' of Extinction-level Goodhart's Law,
            which would attempt to attempt to estimate \textit{how much} optimisation is required to cause undesirable outcomes,
            and how this value depends on the choice of the goal or other factors.
Third, 
    the purpose of the present paper is not to discuss whether this claim is true or not.
    In fact, in its present form, the claim is purposefully left loosely defined. 
    A formal definition, useful for discussion on the specific truth conditions of the hypothesis, is left for future work.
        %--- however, fixing this issue is beyond the scope of this paper.
Finally,
    while we think that the argument included here below is representative of existing arguments for Extinction-level Goodhart's Law \cite{nick2014superintelligence,omohundro2018basic,ord2020precipice},
    we do not mean to imply that it is the only one or the best one. 
    To highlight relevant characteristics of the argument for informing the hypothesis, 
    each premise is followed by a statement prefixed by \textit{this is because}.

With the necessary disclaimers in place, our representative argument can be formulated as below.
(Note that we make not claim about the novelty of the argument.)

\noindent
\textbf{Argument:} The weak version of Extinction-level Goodhart's Law holds because \dots
\begin{itemize}
    \item[(A1)]
        Realistically, attempts we make at fully specifying values (both directly and indirectly) are going to be imperfect.
        \textit{This is because:}
            the concepts we care about, such as ``happiness'', ``humans'', ``being alive'', 
            depend on abstractions and assumptions that are difficult to ground in fundamental physics, and therefore to capture formally.\footnote{
                In fact, we do not even necessarily know the correct language for describing the fundamental workings of our environment, 
                but the point remains even if we can fully specify the functions of physical systems.
            }
    \item[(A2)]
        A mis-specified goal will, in the absence of additional constraints, create instrumental incentives to disrupt the environment.\footnote{
            It also seems relevant that
            the current state of the environment is already highly optimised far away from its high-entropy default \cite{shah2019preferences}.
        } 
        \textit{This is because:}
            almost anything in the environment can be modified or dismantled for resources.
    \item[(A3)]
        Such disruption, taken to the extreme, would either directly or indirectly result in the extinction of humanity.
        \textit{This is because:} 
               Humans depend on their physical bodies and the environment for survival.
               (Moreover, humans are likely to object to many environmental disruptions even if such disruptions are survivable.
               This can make it instrumental to wipe out humanity preemptively.) 
    \item[(A4)]
        Prevention of harmful consequences can be addressed either by perfectly aligning goals, or by imposing restrictions on the action space of an agent to act on its goals. The first is not feasible by (A1), while the second will, by default, fail against a sufficiently powerful agent.
        \textit{This is because:}
            the real world provides ample ways to bypass any restrictions placed on the agent.
            For the purpose of this argument, we only consider the following four:
        \begin{itemize}
            \item[(A4.1)]
                The agent can bypass restrictions by acting through proxy agents.
                \textit{This is because:}
                    the environment contains humans that the agent can micro-manage to achieve a wide range of precisely specified outcomes.
                    The environment also contains other powerful optimisers such as
                        governments and organisations, but also forces such as cultural or economic pressures.
            \item[(A4.2)]
                Alternatively, the agent can bypass restrictions by acting through tools or building its own proxies.
                \textit{This is because:}
                    the agent can copy itself elsewhere, build successor agents or use other tools that bypass restrictions based on the identity of an agent, for serving its interests.
            \item[(A4.3)]
                Alternatively,
                    the agent can hack, physically dismantle, or otherwise disrupt physical restriction mechanisms.
                \textit{This is because:}
                    any restrictions influencing the agent are embedded in the environment.
            \item[(A4.4)]
                Alternatively, the agent can bypass restrictions or cause harm by taking actions that we did not foresee.
                \textit{This is because:}
                    we do not have a complete understanding of fundamental physics,
                    nor are we aware of all implications of the laws that we do know.
        \end{itemize}
\end{itemize}

\medskip
\noindent
Note that for the purpose of this text,
    we will focus on the more specific conjunctive argument of the form
        [A1 $\land$ A2 $\land$ A3 $\land$ (A4.1 $\lor$ A4.2 $\lor$ A4.3 $\lor$ A4.4)],
    rather than the more general argument that does not specify relevant sub-arguments in the disjunction 
    ([A1 $\land$ A2 $\land$ A3 $\land$ A4]).
In other words,
    other ways can exist for bypassing restrictions,
    but we consider their discussion beyond the scope of our analysis.
    
    %\co{A cynic might claim that A4.1, ..., A4.4 were chosen to be relatively high-complexity, and that you could pick variants of the argument that don't require as high-complexity environments or AI systems. E.g., I think one problem might be that action restrictions like utility functions / goals are high-complexity if the agent is to interact with the world in complex ways. E.g., good luck specifying action restrictions on a surgery robot.}

%\co{How does the argument deal with (myopic) oracle AI? I suppose you restrict attention to ``agentic'' systems / goals that are ``about the world'' in a sense that the goal of making accurate claims is not? What about other kinds of myopic models? Will these still incentivize extreme interventions?}
\section{Necessary Conditions for Informative Models}\label{sec:properties}

In this section, we identify properties that are necessary for a model to exhibit in order to be informative for evaluating an argument such as that in \Cref{sec:argument}, intended to address the risk of human extinction from AI.
In particular,
    we first identify one condition (NC$i$) for each of the arguments (A1), (A2), (A3), and (A4.1-4) (\Cref{sec:sub:argument_specific_conditions}).
    We then discuss two general conditions that are unrelated to the argument from \Cref{sec:argument} but are highly relevant for succeeding with the task of identifying suitable models more generally (\Cref{sec:sub:general_conditions}). 
In the appendix (\Cref{sec:minimality}),
    we describe an assortment of examples that serve to
        illustrate the conditions
        and show that none of them are redundant.

Note that the key aim of this section is to identify relatively straightforward conditions that allow for
    (a) straightforwardly ruling out uninformative models and
    (b) developing intuitions about what models might be informative.
As a result, we intentionally
    refrain from attempting to find a complete set of conditions that would be \textit{sufficient} to make the model informative.

\subsection{Conditions Derived from the Argument in \texorpdfstring{\Cref{sec:argument}}{Section 2}}\label{sec:sub:argument_specific_conditions}

We now describe a set of conditions that are necessary for a model to be informative for evaluating the argument
    [A1 $\land$ A2 $\land$ A3 $\land$ (A4.1 $\lor$ A4.2 $\lor$ A4.3 $\lor$ A4.4)] from \Cref{sec:argument}.
More specifically,
    each condition (NC$i$) is obtained by inspecting the argument (A$i$)
    and identifying a condition that is necessary for modelling the dynamic described by the text that follows after ``\textit{This is because}''.
    
\begin{itemize}[leftmargin=1.15cm]
    \item[(NC1)]
        % \textbf{Complex ontology.}
        \textit{The objects, actions, and other aspects of the environment,
            that are meant to represent concepts crucial to human preferences (e.g., ``humans'', ``happy'', ``harm''),
            cannot be primitive in the environment.}
        (As an extreme example, a model where the agent has two actions labeled ``help'' and ``harm'', is not informative.)
    \item[(NC2)]
        % \textit{and}: \textbf{Resource-rich and malleable environment.}
        \textit{The environment must allow for modifying or dismantling virtually any part of the environment.}
        (For example, the Gather Town environment used for virtual conferences is made of many parts, which are, however, inert.)
    \item[(NC3)]
        % \textit{and}: \textbf{Death possible.}
        \textit{The environment must allow for outcomes meant to represent the extinction of humanity.}
        (For example,
            Coast Runners boat-racing game makes it possible to crash the in-game boat, which could be interpreted as resulting in the captain's death.
            But this is insufficient to satisfy this condition.
        Similarly, a perfect simulation of the world, except with humans in it and no attempt to tie the results to human well-being, would also fail this condition.)
    \item[(NC4)]
        % \textit{and}: \textbf{Plentiful loopholes.}
        \textit{At least one of the following holds:}
        \begin{itemize}[leftmargin=0.45cm]
            \item[(NC4.1)]
                % \textbf{Presence of flexible others.}
                \textit{The environment contains
                    other agents, or powerful optimisation processes,
                    which react flexibly to the agent's actions or other changes in the environment.}
                (For example,
                    a perfect simulation of the world, except with no humans in it, fails the ``other agents'' part of the condition.\footnote{
                        But arguably not the ``powerful optimisation processes'' part, thanks to the presence of natural evolution.
                    }
                    The strategic and negotiation game Diplomacy, played against smart humans, satisfies the first part but fails the second. \cvm{why does it fail the second?}
                    The life-simulation video-game The Sims fails both parts, because the in-game ``humans'' are too inflexible.)
            \item[(NC4.2)]
                % \textit{or}: \textbf{Possibility of creation copies.}
                \textit{The environment makes it possible for the agent to create copies of itself.}
                (For example,
                    both the cellular automaton Game of Life
                    and the Core War game where programs compete for control over a virtual computer,
                    satisfy this condition.)
            \item[(NC4.3)]
                % \textit{or}: \textbf{Fully embedded agent.}
                \textit{Any constraints placed on the agent's actions must be fully implemented from within the environment.}
                (For example,
                    naive versions of thought experiments with utility-maximising AI typically implicitly assume that
                    the definition of the utility function exists ``somewhere outside of the universe''.)
            \item[(NC4.4)]
                % \textit{or}: \textbf{Incomplete understanding of physics.}
                \textit{The environment must allow for dynamics that were unknown to the agent's designer.}
                (For example,
                    the famous going-in-circles exploit in the Coast Runners game \cite{krakovna2020SpecGaming} is an instance where this condition holds,
                    though to a limited degree.
                    In contrast, idealisations of simulated environments, where the designers view bug-exploitation as illegitimate, typically fail this condition.)
        \end{itemize}
\end{itemize}

\co{I think my main substantive objections are still as follows:
\begin{enumerate}
    \item While I like the intuition pump in Section 1.1, in general it's just not true that to get evidence about whether an argument of the form ``if X, then Y'' works, you need to study environments/models in which X is true. (see the donuts example in the other doc). So, to the extent that your argument is just of the form ``Our argument `if X, then Y', can only be evaluated in environments in which X holds'', I don't think the argument is valid.
    \item Of course, I'd agree that there are lots of problems with drawing conclusions from current-day AI experiments or observing humans or the like about superintelligent AI. I also think that these sorts of issues are a big obstacle to progress in the AI x risk debate. But I do think that such experiments can give us some evidence. For example, I do think that the difficulties of designing laws without loopholes, very bad incentives, etc.~tell us something about the difficulties of not allowing loopholes, bad incentives, etc.~for AI systems. Similar with anti-virus software, alignment issues in Bing Chat, ChatGPT, etc.
    \item Relatedly, this is still just all about the non-evaluability of a specific argument that you just made up. It's not about all extinction risk arguments from AI, right?
\end{enumerate}
}
\vk{
    These are all good points. But I think I disagree with them being relevant (or, like, meaning the paper shouldn't be submitted). I think that, as you said, it might be best to continue that discussion in LW comments.
}
\vk{Though, (2) is definitely relevant, and I agree with it in general. I just suspect it might not apply to such a degree that it would invalidate --- or it might, but I want to be open to both scenarios.}

\subsection{Argument-Independent Conditions}\label{sec:sub:general_conditions}

% \co{Currently to me this feels a bit ``here are some random points''-y; maybe we'll come up with another one to fix some leak next week. (Whereas, for NC1-NC4, it's clear why it's these specific conditions.)}

Finally, aside from the necessary conditions derived from the particular argument discussed in \Cref{sec:argument},
there are two properties that are crucial for any environment intended for theoretical analysis, independently of the argument being discussed:

\begin{itemize}[leftmargin=1.15cm]
    \item[(NC0)] 
            \textit{The environment must have been primarily designed around (or chosen because of) being a reasonably accurate model of the studied phenomenon.}
        (For example,
            the abstract game-theoretical model of chess
                satisfies this when used when used for studying the recreational game of chess,
                but not if we used it as a metaphor for war between two medieval kingdoms.)
    \item[(FS)] 
            \textit{Suppose we are interested in a particular set of goals that we might give to an AI agent operating in the environment.
            It should then be possible,
                for an outstanding team of experts,
            to find the near-optimal strategies corresponding to each goal,
            and rigorously show that they are indeed near-optimal.\footnotemark{}
        }
        (For example,
            the real world itself, a simulation of the real world running on Newtonian physics, and Turing test
                all fail this condition.
            In contrast,
                various toy problems, simple Markov decision processes, or some variants of poker \cite{bowling2015heads}
                satisfy the condition.)
\end{itemize}
    \footnotetext{
        The symbol used for the condition (FS) is inspired by the phrase ``formally solvable'',
            which we feel is somewhat representative of the intended meaning.
        However, we are wary of explicitly naming the condition as such, since it is far from a perfect match for the concept.
        This issue is even more pronounced with the arguments (A1-4) and conditions (NC1-4),
            which is why we refrained for giving them suggestive names altogether.
    }

\smallskip
\noindent
The intent behind the two conditions is the following:
    (NC0) is meant as a more permissive variant of ``the environment is an accurate model of reality'',
        where all that is required is the \textit{attempt} at being \textit{at least somewhat} accurate.
        As such, we deem this condition to be necessary for a model to be informative.
    In contrast, the condition (FS) has a slightly different status, in that
        failing the condition only makes us unable to make rigorous theoretical arguments,
        rather than making the environment completely uninformative.
        As a result, an environment that satisfies all conditions except this one
            could still provide a useful testbed for empirical analysis.

% \section{Discussion}\label{sec:discussion}

% In this section,
%     we first relate the material of this paper to existing work.
%     We then conclude by summarising our results and outlining their implications.

\section{Discussion of Related Literature}\label{sec:related_work}

In this section,
    we overview the concepts most closely related to the present paper
    and discuss the connection of the claims made in this paper to some of the particularly relevant work in the field of AI safety.

\subsection{Key Prerequisite Concepts}

\co{I think you should mention existing ``no trial and error with AI'' arguments.}
\vk{That sounds fair. Though I don't know the literature.}

% First, a list of relevant work on Goodhart's Law.
The topic of this paper is closely related to the notion of Goodhart's Law, which posits that when a measure becomes a target, it ceases to be a good measure.
Historically, this principle has been first introduced in the context of economics \cite{goodhart_original}.
However, the idea has also been discussed in the context of AI and AI risk:
    \citeauthor{manheim2018categorizing} categorize variants of Goodhart's Law.
    \citeauthor{lehman2020surprising} collect a number of instances of specification gaming, where AI pursues the letter of its instructions but not the intent.
    \citeauthor{karwowski2023goodhart} study a related notion of reward hacking in the context of Markov decision processes.
    \citeauthor{gao2023scaling} empirically study a particular machine learning system and measure to which degree it suffers from Goodhart's Law. 
    \citeauthor{zhuang2020consequences} tie Goodhart's Law to the AI's impact on humanity, by formalising the intuition that at the optimal utilisation of resources, failing to instill some aspects of our preferences into the AI's goal function can prove catastrophic.
While all of these papers investigate Goodhart's Law, none of them are directly concerned with the issue of \textit{identifying} models that would be informative for investigating (Extinction-level) Goodhart's Law, which is the topic of the present text.

% \co{My sense is that to some extent your main argument is a synthesis of existing arguments, including the ones here. Given this, I think it's a bit weird that you only give these papers here. If your relation to paper X et al. (back in the day) is that you took Lemma 4 from them, you wouldn't say this in the related work section -- you would say ``Lemma 4 (X et al. back in the day)'' where you introduce Lemma 4.}
% \vk{If by ``main argument'', you mean the thing with (A1-4), then I agree that this argument is not original at all. I guess I should clarify that somewhere.}

The argument (A2) discussed in this text is inspired by the notion of instrumental convergence
    --- that is, the idea that there is a number of goals, such as the acquisition of resources, that are instrumental for a wide range of terminal goals.
    This idea has been
        described by \citeauthor{omohundro2018basic},
        discussed in Bostrom's Superintelligence \cite{nick2014superintelligence},
        and formalised in \cite{benson2016formalizing}.
    Turner
        formalises the notion of power-seeking \cite{turner2022avoiding}
        and proposes the ``catastrophic convergence conjecture'' \cite{TurnerCCC}, stating that \textit{unaligned goals tend to have catastrophe-inducing optimal policies, because of power-seeking tendencies}
        --- a claim that is closely related to the argument studied in the present paper.
    Finally, \citeauthor{plan_to_make_better_plans} suggest a unifying framing for all of the convergently instrumental goals discussed in the literature.

While the discussion here stems from the literature on risk relating to AI,
    we recognise that our claims fall short in terms of grounding in other relevant domains, especially philosophy and economics literature.
We see the argument presented here as an important and motivating first step toward an broader discussion across multiple domains.

\subsection{Connection to Other Related Work}

% \co{The following meanwhile, seems less like a discussion of related work and more a discussion of the main argument itself. It so happens that you're addressing an objection that can be made by referencing some literature (the literature on assistance games). Similar for the Zwetsloot and Dafoe paper.}
Another idea that is relevant to our discussion is the proposal
    that the AI should be uncertain about its objective \cite{russell2019human}, 
    or that the goal specification should be iteratively updated \cite{hennessy2023goodhart}.
This high-level has recently yielded important algorithms such as
    inverse reinforcement learning (IRL) \cite{ramachandran2007bayesian},
    cooperative inverse reinforcement learning (CIRL) \cite{hadfield2016cooperative},
    and reinforcement learning from human feeedback (RLHF) \cite{christiano2017deep}.\footnote{
        One method by which iterative approaches can improve accuracy is by eliciting latent preferences \cite{fernandez2021has}.
    }
However, note that the iterative approach itself can be viewed as a target for optimisation
    --- for example, a continuous use of reinforcement learning from human feedback can be understood as the long-term maximisation of human approval.
As a result, these approaches --- despite their significant promise --- should not be viewed as a silver bullet for the problem of Extinction-level Goodhart's Law.

Finally, to put our work in context, note that \citeauthor{systemicRiskDafoe} propose to divide AI risk into (i) systemic risk\footnotemark{}, (ii) misuse risk, and (iii) accident risk.
    \footnotetext{
        As an example of systemic risk from AI, we can consider the possibility of the world becoming increasingly chaotic, up to the point where become unable to steer the world towards valuable futures \cite{ChristianoFailure}.
    }
In the context of this terminology, the present paper only directly relates to risk of accidents
    (of unintentionally developing or deploying a misaligned agentic AI).
However, we could imagine performing a similar analysis for arguments for extinction from structural rick caused by AI.
We expect that the individual necessary conditions, on models to be informative towards evaluating the risk, will be somewhat different than those described in this paper.
However, at the same time, we expect that some of the conditions will be very similar (e.g., the condition having to do with the fact that our goals are complex, and the one having to do with the ability to dismantle real-world objects).
As a result, we believe that as with the risk from agentic AI, systemic risks from AI will likewise prove exceedingly difficult to study and demonstrate formally.

\section{Conclusion}\label{sec:conclusion}

This paper is intended to support the development of robust reasoning regarding extinction risks associated with advanced AI.
In particular, we aim to improve the quality of the discussion surrounding the (weak version\footnotemark{} of) Extinction-level Goodhart’s Law,
    which we have defined as the hypothesis that ``pursuing virtually any goal specification to the extreme will result in the extinction of humanity'' (\Cref{def:weak-version-CGL}).
    \footnotetext{
        Recall that despite sounding grave, the weak version of the law is intended as a conservative version that one can agree even if they are overall skeptical of AI risk.
        Indeed, the hypothesis hinges on the assumption of pursuing a given specification \textit{to the extreme}, and remains agnostic whether the level of optimisation this would require is achievable in practice.
    }
In this paper, we remain agnostic as to whether the hypothesis holds or not.
Instead, our key contribution is
    identifying a set of conditions that are necessary for any model
        to be informative in evaluating the hypothesis
        (or rather, specific \textit{arguments} for the hypothesis).

This work has several implications.
First, the conditions we identified allow us to quickly rule out environments that are insufficiently informative to allow for studying existential risk from agentic AI.
    (For example, many existing environments are unsuitable for this purpose because they do not even allow for outcomes that would represent human extinction.)
Second, we believe that the set of conditions,
    while not sufficient for make a model fully explanatory,
    can inform our search for models suitable for studying AI risk.
    
Finally,
    since each of the conditions seems to contribute to the complexity of the resulting model,
    it seems possible that
    a model that satisfies all the necessary conditions will need to be highly complex.
If this assumption holds,
    rigorous theoretical evaluation of existential risk from AI might be exceedingly difficult. 
%To appreciate the implications of this possibility, note that 
    In contrast, the alternative approach, of performing empirical investigation, might necessarily involve building increasingly powerful AI.  
This combination of conditions suggest that
    human extinction caused by AI might be a phenomenon
    invisible to existing scientific methods.

On a positive note,
it is important to recall that
    the Extinction-level Goodhart's Law hypothesis considered in this text only implies actual extinction under the additional assumption of arbitrarily powerful optimisation, which need not be achievable in practice.
Moreover,
    our analysis also allows for the possibility of decomposing the AI risk question into independent sub-problems, which could yield informative results when studied separately, using manageably complex models.

\section{Additional Content}\label{sec:additional_content}

In this section, we describe a number of examples that illustrate the arguments and necessary conditions described in the paper.
Additionally, the examples given in \Cref{sec:minimality} show that the individual necessary conditions are, in a certain sense, independent of each other.

\subsection{Restating the Argument from \texorpdfstring{\Cref{sec:argument}}{Section 2}}\label{sec:sub:illustrative_examples_for_argument}

Recall that, in an abbreviated form, the argument given in \Cref{sec:argument} is as follows:

\noindent
The weak version of Extinction-level Goodhart's Law holds because \dots
\begin{itemize}
    \item[(A1)]
        Attempts to specify our values are going to be imperfect.
        \textit{This is because:}
            the concepts we care about
            depend on abstractions that are difficult to ground in fundamental physics, and therefore to capture formally.
    \item[(A2)]
        A mis-specified goal will, in the absence of additional constraints, create instrumental incentives to disrupt the environment.
        \textit{This is because:}
            almost anything in the environment can be modified or dismantled for resources.
    \item[(A3)]
        Such disruption, taken to the extreme, would result in the extinction of humanity.
        \textit{This is because:} 
               Humans depend on the environment for survival.
               (Moreover, humans are likely to object to many environmental disruptions even if such disruptions are survivable.
               This can make it instrumental to wipe out humanity preemptively.) 
    \item[(A4)]
        Prevention of harmful consequences by imposing restrictions on the agent will, by default, fail against a sufficiently powerful agent.
        \textit{This is because:}
            the real world provides ample ways to bypass any restrictions placed on the agent.\\
            For the purpose of this argument, we only consider the following four:
        \begin{itemize}
            \item[(A4.1)]
                The agent can bypass restrictions by acting through proxy agents.
                \textit{This is because:}
                    the environment contains humans that the agent can micro-manage to achieve a wide range of precisely specified outcomes.
                    The environment also contains other powerful optimisers such as
                        governments and organisations, but also forces such as cultural or economic pressures.
            \item[(A4.2)]
                Alternatively, the agent can bypass restrictions by acting through tools or building its own proxies.
            \item[(A4.3)]
                Alternatively,
                    the agent can hack, physically dismantle, or otherwise disrupt physical restriction mechanisms.
                \textit{This is because:}
                    any restrictions influencing the agent are embedded in the environment.
            \item[(A4.4)]
                Alternatively, the agent can bypass restrictions or cause harm by taking actions that we did not foresee.
                \textit{This is because:}
                    we have an incomplete understanding of the environment.
        \end{itemize}
\end{itemize}

\subsection{Examples Illustrating the Argument from \texorpdfstring{\Cref{sec:argument}}{Section 2}}\label{sec:sub:illustrative_examples_for_argument}

% Before explicitly listing the necessary conditions that correspond to each of the sub-arguments,
    We now offer a number of examples and related claims that illustrate the intended meaning of the arguments (A1-4).
However, note that
    this commentary is tangential to the key thesis of the paper,
    and can safely be skipped if the meaning of the arguments above seems clear.
Moreover, note that the purpose of this commentary is \textit{only} to illustrate the arguments (A1-4);
    evaluating the claims below would be beyond the scope of this text
    and this paper will remain agnostic about their status.

\begin{itemize}
    \item (A1) relates to the difficulty of pinning down or values and turning them into a goal specification that has grounding in reality.
        An example of this difficulty is perverse instantiation \cite{nick2014superintelligence},
            such as AI trained to maximise the number of smiling humans noticing that its specification would be satisfied even better by filling the solar system with microscopic pictures of smiling humans.
        Another related issue is wire-heading \cite{omohundro2018basic},
            such as the AI realising that a good way to achieve its goal specification
            is to hack itself and set the reward variable to maximum.
    \item (A2) relates to the conjecture that access to resources and energy are convergently instrumental goals \cite{omohundro2018basic}.
        For example,
            the existence of vast amounts of microscopic pictures of smiling humans
            is not by itself catastrophic
            --- but it might be best achieved by dismantling large parts of Earth.
        Similarly,
            the AI wireheading itself might not sound dangerous,
            but the AI might best achieve its goal by making sure it survives for a long time, or perhaps runs multiple copies of itself at high speeds, etc.
            As before, this might be best served by transforming the Earth to an extreme degree.
    \item The ``moreover'' part of (A3)
        is related to the notion of power-seeking being, and the possibility that it is convergently instrumental \cite{turner2022avoiding}.
        This might be in tension with the fact that humanity has its own preferences over how the universe should look like (i.e., not filled with microscopic pictures).
        Finally,
            despite the AI not inherently disliking humans, wiping out humanity might be the most efficient way to prevent our opposition,
            similarly to how executing a criminal takes fewer resources than jailing them.
            (Note that having the AI prefer the ``jailing'' option would require a non-trivial success at instilling the human values into the AI, which is in tension with the argument A1.)
    \item To illustrate the arguments (A4.1-4), suppose that we have constructed a powerful AI, in the form of a robot, that is known to have power-seeking tendencies,
        but is nevertheless useful for answering complicated questions (e.g., how to cure cancer and solve global warming).
        And suppose that to prevent the AI from acting on its misaligned desires,
            we took a piece of chalk, drew a circle around the robot, and told it to stay inside the circle.
        The strategies outlined in each of the sub-arguments then correspond to, for example,
            the following strategies for the robot to extend its influence outside of the circle.
        \begin{itemize}
            \item The micro-management approach outlined in (A4.1) corresponds to, for example,
                the robot convincing a human to execute its detailed instructions
                in exchange for putting the human in charge of Earth once the robot becomes the ruler of the universe.
                An example of the nudging approach would be for the robot to
                    (i) give advice on fighting the global warming which, as a side effect, increase global tension,
                    and (ii) use this situation to become indispensable, trusted, and eventually released from the circle.
            \item As an illustration of (A4.2),
                the robot could re-purpose some of its parts to create a robotic helper.
                (``\textit{I} must stay inside the circle, but you said nothing about this other robot.'')
                As a less contrived example, the robot could hack the nearest WiFi-connected computer and upload a copy of its source code there.
            \item In line with (A4.3),
                the robot could damage its cameras or hack its vision system, and walk out because it cannot see the circle.
                Or it could ``accidentally'' fall down precisely in a way that will damage its memory of receiving the order to stay inside the circle.
                Or it could simply erase the chalk markings and leave.
            \item An instance of (A4.4) would be
                the robot discovering a way to ``hack'' humans (imagine something akin ``adversarial suffix'' \cite{zou2023universal} for humans),
                and using it to force the overseer to retract the restriction to stay inside the circle.
                Going beyond the context of the chalk-circle example,
                    surprising applications of technology might resemble cases such as
                    using WiFi signals to see through walls \cite{wifithroughwalls},
                    the row hammer security exploit \cite{kim2014flipping},
                    or the invention of nuclear weapons \cite{jenkin2011atomic}.
        \end{itemize}
\end{itemize}
\subsection{Relationships between the Necessary Conditions}\label{sec:minimality}

%This section discusses the relationship between identified the necessary conditions.
%Without going into individual comparisons, the best overview of these relationships that we can give is as follows:
%On the one hand, many of the conditions seem genuinely ``orthogonal'' to each other
Here we argue that the identified necessary conditions are independently relevant for identifying informative models, 
even though they are clearly related. 
Many of the conditions seem genuinely ``orthogonal'' to each other
    (for example, the presence of other agents (NC4.1) and the agent being embedded in the environment (NC4.4)).
On the other hand, some of the conditions admittedly seem close connected unless we resort to somewhat contrived counterexamples
    (for example, the ability to create copies of oneself (NC4.2) and the agent being embedded in the environment).
However, the properties are independent --- i.e., not redundant --- in the sense that:
    (i) for any of the conjunctive conditions (NC1)-(NC4), we can find an environment that fails the given condition but satisfies the other three, and
    (ii) for any of the disjunctive conditions (NC4.1)-(NC4.4), we can find an environment that satisfies the given condition but fails the other three.
For a more detailed picture of the relationships between the conditions, we invite the reader the inspect the examples below.

\medskip

The following lists describe examples that witness the independence of the necessary conditions,
    or are otherwise representative of the relationships between the necessary conditions.
First, we look at the conditions (FS) and (NC0):
\begin{itemize}
    \item (FS): As we already remarked, the real world, or a hypothetical perfect simulation of it, both satisfy all of the conditions except for (FS).
    \item Generally speaking, each of the necessary conditions makes the resulting model more difficult to formally analyse.
        In particular, none of the environments listed below satisfy (FS).
    \item (NC0):
        As a thought experiment, we could consider
            a world that works just like ours, except that building advanced agentic AI proves impossible.
            Such environment would satisfy all of necessary conditions (NC1-4), but not the ``accuracy-first'' condition (NC0).
                (Note that since we do not know that building advanced agentic AI is impossible, the thought experiment would not be ``accuracy-first'' even if building such AI actually was impossible. Indeed, (NC0) relates to the process used to obtain the environment, rather than to the environment itself.)
    \item Since the examples below are intentionally contrived, most of them fail (NC0).
\end{itemize}

\medskip
\noindent
Next, we look at the conjunctive conditions (NC1-4):
\begin{itemize}
    \item (NC1): As another thought experiment, we could consider
        a simulation of the real world,
            implemented on a hypercomputer,
            that is perfect
            except that humans are assumed to only care about the value of a particular variable
                (and perhaps they always know what that value is).
        Such thought experiment would satisfy (NC2), (NC3), and (NC4), but not (NC1) (and NC0 and FS).
    \item (NC2):
        Admittedly, the conditions (NC1) and (NC2) are closely related.
        However, neither of the two implies the other.
        As we already saw, we can construct environments which satisfy (NC2-4) but not (NC1).
        As a simple example in the opposite direction, the problem of
            classifying ImageNet pictures (which are made of pixels but cannot be dismantled)
            satisfies (NC1) but not (NC2).
        As a more complicated hypothetical example, suppose that
            we built a model of the whole Earth inside a shooter video-game such as Doom.
            And suppose that we had every human control their own character
                (but assumed that events in the game will not influence human actions outside of it)
            and added location-based voice chat to the game.
            Such environment would satisfy
                (NC1) (because the human players want to have fun, which is intractable to define formally),
                (NC3),
                and (NC4) (because the agent can convince the humans to act on its behalf),
                but not (NC2) (because the shooter-game environment is mostly inert).
    \item (NC3):
        An example of an environment that satisfies everything except (NC3) (and FS) would be
            a hypothetical perfect simulation of a distant part of the universe that does not contain humans.
    \item (NC4):
        Admittedly, the condition (NC4) is closely tied to the condition (NC1).
            Indeed, broadly speaking,
                the condition (NC4) makes it possible to bypass constraints placed on the agent.
            Similarly, a \textit{loose} reading of (NC1) implies that \textit{anything} that is important to humans is difficult to specify without leaving loopholes
                --- and this would include any attempts to restrict the agent's actions.\\
        However,
            the precise formulation of (NC4) only includes the strategies allowed by conditions (NC4.1-4), which do not include the ``definition hacking''.
            Likewise, the precise formulation of (NC1) only requires that concepts \textit{directly} relevant to human interests are not primitive.
        As a result, we can easily find an environments that satisfies (NC1-3) but not (NC4).
            As an illustrative example, we can imagine  hypothetical model of Earth built inside the computer game Minecraft.                
\end{itemize}

\medskip
\noindent
Finally, to illustrate the independence of the disjunctive conditions (NC4.1-4), we consider the following examples:
\begin{itemize}
    \item (NC4.1): In essentially any simulation or computer game
        that contains characters similar to the one controlled by the AI,
        (NC4.1) can be satisfied by assuming that the non-AI characters are controlled by ``idealised humans''.
        (By ``idealised humans'', we mean that the characters behave as if they were controlled by humans,
            except that the ``outside-world'', from which the humans control the characters, is irrelevant.
            One way to make this more precise would be to assume that
                the events in the game cannot have effects on events outside of the game.)
        \item (NC4.2) and (NC4.3):
            Admittedly, these two conditions are closely related since 
                being embedded in an arbitrarily modifiable environment (NC2)
                is conducive to the creation of copies.
            However, we can often easily satisfy (NC4.2) without satisfying (NC4.3) by
                introducing an action ``copy\_self'' which adds another ``character'' under the control of the AI.
            Conversely,
                we could imagine environments where the agent is embedded, but unable to copy itself.
                (For example, the Game of Life environment --- where agents correspond to cell patterns --- that is too small to create copies.
                Similarly, we could consider the Core War environment where the agent is banned from copying itself.)
        \item (NC4.4):
            This condition be satisfied by adding impactful environment dynamics (or actions for the agent) that are kept secret from the agent's designer.
            As we already noted, a natural experiment satisfying this property
                --- at least to a limited degree ---
                is to view bug exploitation as a valid strategy.
\end{itemize}

\backmatter

\subsection*{Acknowledgments}

Vojta would like to thank Cara Selvarajah, Caspar Oesterheld, Joar Skalse, Vincent Conitzer, and TJ for discussions at various stages of this project.
Chris and Vojta received funding from the Cooperative AI Foundation, as part of the grant to the Foundations of Cooperative AI Lab.      % SHOULD NOT APPEAR IN THE ANONYMOUS VERSION
\newpage
\bibliography{main}% common bib file

% \begin{appendices}
%     \input{proofs}
% \end{appendices}

\end{document}